\documentclass[onecolumn,showpacs,preprintnumbers,amsmath,amssymb]{revtex4}

\usepackage{amsmath,color}
\usepackage{graphicx}
\usepackage{dcolumn}
\usepackage{bm}

\begin{document}


\title{A route to quasi-perfect invisibility cylindrical cloaks without extreme values in the parameters}

\author{Sheng Xi}
\author{Hongsheng Chen}
\email{chenhs@ewt.mit.edu}
\author{Baile Zhang}
\author{Bae-Ian Wu}
\author{Jin Au Kong}
\affiliation{Research Laboratory of Electronics, Massachusetts Institute of Technology, Cambridge, MA 02139, USA, and The Electromagnetics Academy at Zhejiang University, Zhejiang University, Hangzhou 310027, China}


\begin{abstract}
The method of coordinate transformation offers a way to realize
perfect cloaks, but provides less ability to characterize the
performance of a multilayered cloak in practice. Here, we propose an
analytical model to predict the performance of a multilayered
cylindrical cloak, based on which, the cloak in practice can be
optimized to diminish the intrinsic scatterings caused by
discretization and simplification. Extremely low scattering or
``quasi-perfect invisibility" can be achieved with only a few layers
of anisotropic metamaterials without following the transformation
method. Meanwhile, the permittivity and permeability parameters of
the layers are relatively small, which is a remarkable advantage of
our approach.
\end{abstract}

\maketitle

Various efforts have been made on the realization of
invisibility\cite{pendry2006science,greenleaf2003pm,engheta2005pre,engheta2008prl,leonhardt2006science}.
Pendry \emph{et al.} theoretically proposed the perfect invisibility
cloak for electromagnetic waves\cite{pendry2006science}, utilizing
anisotropic and inhomogeneous media to mimic the space squeezing.
Later, the effectiveness of the transformation based cloak was
demonstrated by ray tracing\cite{schurig2006oe}, full wave finite
element simulations\cite{cummer2006pre,zolla2007ol}, and analytical
scattering
models\cite{hongsheng2007prl,baile2007prb,ruan2007prl,baile2008prl},
as well. In practice, the difficulty in construction a perfect
invisibility cylindrical cloak is the requirement of continuous
inhomogeneity and high anisotropy with extreme values in the
parameters. Simplified parameters based on the coordinate
transformation were then utilized to facilitate the physical
realization\cite{cummer2006pre,schurig2006science,cai2007}, in
expense of the aroused inherent scatterings\cite{yanmin2007prl}.
Constraints on the bandwidth were studied as well\cite{chan2007prb}.
The first sample of cylindrical cloak has been created using
multilayered metamaterials \cite{schurig2006science}. Bi-layered
isotropic media was also proposed for achieving the effective
anisotropy\cite{fengyj2007oe}, but a lot of thin layers are needed
which increases the construction complexity. Moreover, the
transformation method provide less ability to predict the
performance of a practical construction composed of discontinuous
layers of homogeneous anisotropic metamaterials. Therefore, it is
very necessary to investigate a better way to design a practical
cloak with good performance.

In this paper, in order to get the exact behavior of a multilayered
cloak, the analytical model of a cylindrical cloak created with
multilayered anisotropic materials is established based on the full
wave scattering theory\cite{Roth:73,chewbook}. Our results show
that, by using only a few layers of anisotropic materials, a
``quasi-perfect invisibility" multilayered cloak with near zero
scattering can still be achieved without following the design method
of coordinate transformation, and the parameters obtained are
relatively small and possible to be realized by metamaterials. The
impedances between the adjoined layers do not really match each
other but can produce zero reflection, which can be treated as the
counterpart in cylindrical geometry of the reflectionless
one-dimensional multilayered slab. All of these results provide a
second and better way of designing a multilayered cloak.
\begin{figure}
\includegraphics[width=0.5\columnwidth,draft=false]{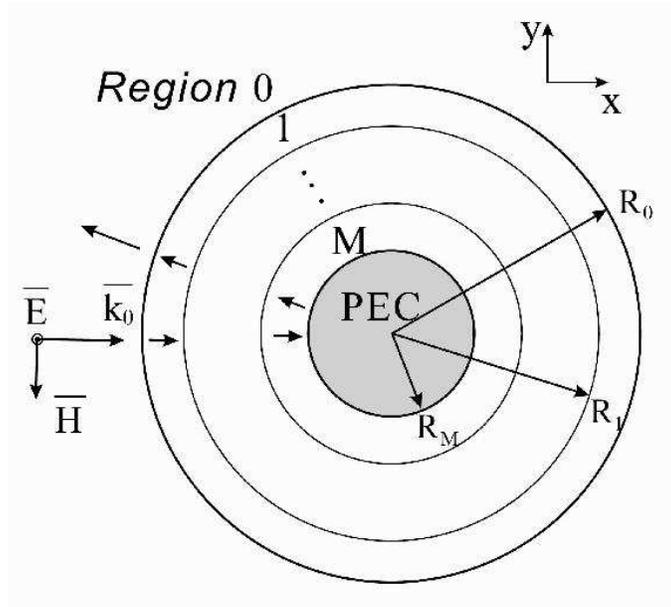}
\caption{\label{fig:configuration} Configuration of a multilayered
cylindrical cloak.}
\end{figure}

We use cylindrical cloak as an example. Without losing the
generality, the case a TE-polarized plane wave with unit magnitude
normally incident onto an $M$-layer cylindrical cloak (from region
$1$ to $M$) is considered, as shown in Fig.~\ref{fig:configuration}.
The TM case can be analyzed similarly. The radiuses of the
boundaries of the cloak are denoted by $R_m$ $(m=0,1,\cdots,M)$. The
relative constitutive parameters in region $m$ are assumed to be
constants denoted by $\mu_{\rho m}$, $\mu_{\phi m}$ and $\epsilon_{z
m}$ while the region $m=0$ is assumed to be free space and the core
region ($\rho<R_M$) is assumed to be PEC. The electric fields
$E_{zm}$ in the region $m$ satisfy the following equation,

\begin{eqnarray}
\frac{1}{\rho}\frac{\partial}{\partial\rho}\left(\frac{\rho}{\mu_{\phi
m}}\frac{\partial
E_{zm}}{\partial\rho}\right)+\frac{1}{\rho^2}\frac{\partial}{\partial\phi}\left(\frac{1}{\mu_{\rho
m}}\frac{\partial E_{zm}}{\partial\phi}\right)\nonumber
\\+k_{0}^2\epsilon_{z m} E_{zm}=0.
\end{eqnarray}
By applying the method of separation of variables, the general
expression for the electric fields in region $m$ can be expressed as
\begin{eqnarray}
E_{zm}=\sum_{n=-\infty}^\infty a_{mn}\left(J_{\nu_{mn}}(k_m
\rho)+\tilde{r}_{m(m+1)n}H_{\nu_{mn}}(k_m \rho)\right)\nonumber
\\\exp(in\phi),\label{eq:Efield}
\end{eqnarray}
where $\nu_{mn}=n\sqrt{\mu_{\phi m}/\mu_{\rho m}}$ and the wave
number in region $m$ is $k_m=k_0 \sqrt{\epsilon_{zm}\mu_{\phi m}}$.
Different from the isotropic layered case, here, $\nu_{mn}$ is a
fraction. $J_{\nu_{mn}}$, $H_{\nu_{mn}}$ represent the $\nu_{mn}$th
order Bessel functions of the first kind and the $\nu_{mn}$th order
Hankel functions of the first kind, respectively. $a_{mn}$ is the
unknown coefficients and $\tilde{r}_{m(m+1)n}$ is the scattering
coefficient on the boundary $R_m$. When a standing wave incident
from region $m$ onto the the boundary $R_m$, the direct reflection
coefficient, which represents the ratio between the directly
reflected wave and the incident wave, is
$r_{m(m+1)n}=(j'j_1-{\eta_m}/{\eta_{m+1}}jj_1')/(-h'j_1+{\eta_m}/{\eta_{m+1}}hj_1')$,
and the direct transmission coefficient, which represents the ratio
between the directly transmitted wave in region $m+1$ and the
incident wave in region $m$, is $t_{m(m+1)n}=-2i/(\pi k_m
R_m)/(-h'j_1+{\eta_m}/{\eta_{m+1}}hj_1')$. Similarly, for an
outgoing wave the direct reflection and transmission coefficients on
$R_m$ are
$r_{(m+1)mn}=(h_1'h-{\eta_{m+1}}/{\eta_m}h_1h')/(-j_1'h+{\eta_{m+1}}/{\eta_m}j_1h')$
and $t_{(m+1)mn}=2i/(\pi k_{m+1}
R_m)/(-j_1'h+{\eta_{m+1}}/{\eta_m}j_1h')$. Here $j=J_{v_{mn}}(k_m
R_m)$, $j'=J_{v_{mn}}^{'}(k_m R_m)$, $j_1=J_{v_{(m+1)n}}(k_{m+1}
R_m)$, $j_1'=J_{v_{(m+1)n}}^{'}(k_{m+1} R_m)$, $h=H_{v_{mn}}(k_m
R_m)$, $h'=H_{v_{mn}}^{'}(k_m R_m)$, $h_1=H_{v_{(m+1)n}}(k_{m+1}
R_m)$, $h_1'=H_{v_{(m+1)n}}^{'}(k_{m+1} R_m)$, and
$\eta_m=\sqrt{\mu_{\phi m}/\epsilon_{zm}}$ \cite{chewbook}.
Therefore, the scattering coefficient in layer $m$ (m=0,1,...,M-1)
can be written as \cite{chewbook}
\begin{eqnarray}
\tilde{r}_{m(m+1)n}=r_{m(m+1)n}+\tilde{t}_{(m+1)mn}, \label{eq:rto}
\end{eqnarray}
where
$\tilde{t}_{(m+1)mn}={t_{m(m+1)n}t_{(m+1)mn}\tilde{r}_{(m+1)(m+2)n}}/(1-r_{(m+1)mn}\tilde{r}_{(m+1)(m+2)n})$
represents the wave coming out from $R_m$ due to the multiple
reflections and transmissions on the boundaries inside $R_m$. At
$R_M$, $\tilde{r}_{M(M+1)n}=-J_{\nu_{Mn}}(k_M R_M)/H_{\nu_{Mn}}(k_M
R_M)$, therefore all the $\tilde{r}_{m(m+1)n}$ can be derived using
Eq. \ref{eq:rto}, and $a_{mn}$ can also be derived by matching the
boundary conditions. The coefficients of the scattering fields in
region 0 are $b_{0n}=a_{0n}\tilde{r}_{01n}$. Based on the
cylindrical scattering model, the far-field total scattering
efficiency or the scattering cross section normalized by the
geometrical cross section for the multilayered cylindrical cloak is
obtained as
\begin{eqnarray}
Q_{sca}=2/(k_0
R_M)\sum_{n=-\infty}^{\infty}\left|b_{0n}\right|^2.\label{eq:Qsca}
\end{eqnarray}

\begin{figure}
\includegraphics[width=0.5\columnwidth,draft=false]{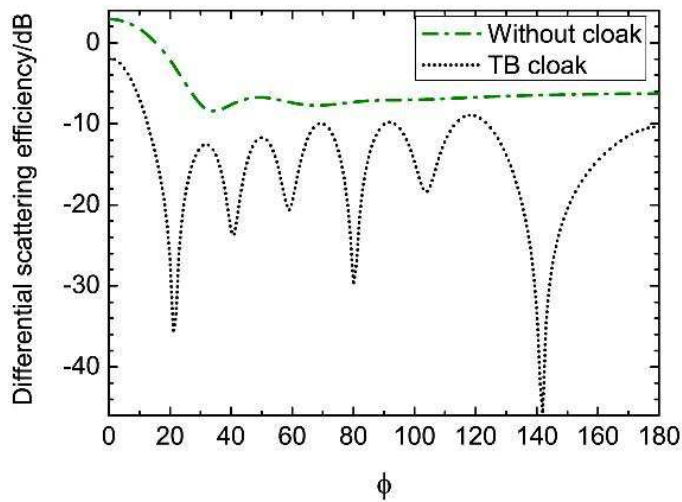}
\caption{\label{fig:smithRCS} Comparison of the differential
scattering efficiencies when a TE-polarized plane wave normally
incident on to the PEC cylinder without cloak (dash dotted line) and
with the cloak proposed in Ref. \cite{schurig2006science} (dotted
line).}
\end{figure}

\begin{figure}
\includegraphics[width=0.5\columnwidth,draft=false]{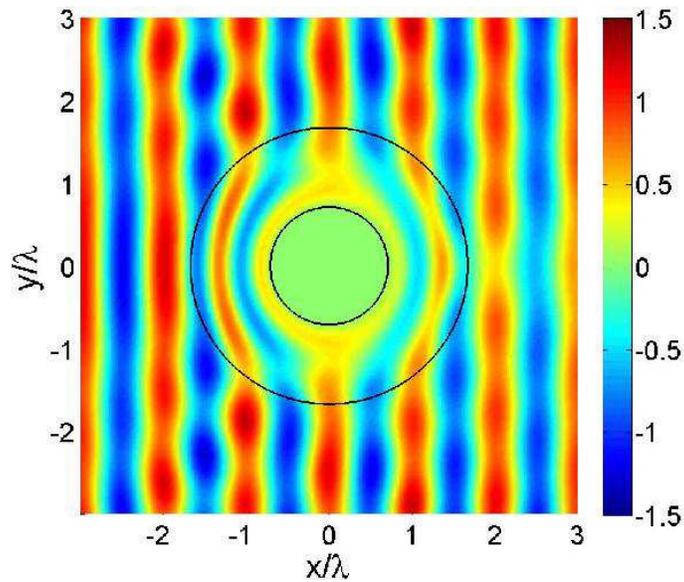}
\caption{\label{fig:smithfields} Electric field distributions for a
plane wave incident from left to right onto the cloak proposed in
Ref. \cite{schurig2006science}.}
\end{figure}

In practice, the ideal parameters obtained from the transformation
method need to be discretized for realization, which will destroy
the perfect invisibility of the cloak. Using the proposed method,
such effects of descretization and simplification of the
transformation based (TB) cloak can be quantitatively analyzed. For
example, Ref. \cite{schurig2006science} proposed a 10-layer
simplified cloak with $\mu_\phi=1$ for the experiment, in which the
copper cylinder core with radius $0.709\lambda$ is coated by a
multilayered cloak with inner radius $0.768\lambda$ and outer radius
$1.670\lambda$. Utilizing the parameters including the losses in the
metamaterials in Ref.  \cite{schurig2006science}, and assuming the
core to be PEC for convenience, for the normal incidence of a
TE-polarized plane wave, the forward and backward scatterings have
been reduced by about 4.8dB and 4.1dB, respectively, comparing with
the bared PEC core, as shown by the dotted line in
Fig.~\ref{fig:smithRCS}.
The near field distributions can also be calculated using our
method. Fig. \ref{fig:smithfields} shows the electric field
distributions of the above case. Our analytical model shows some
qualitative agreement with the experimental field distributions
shown in \cite{schurig2006science}, where both reduced forward and
backward scatterings can be observed.

A more interesting thing is that by using the genetic algorithms, a
widely used optimization method in engineering and science that
enable the individuals of an optimization problem evolves to better
solutions \cite{coleybook}, our proposed method can realize a
general cloak without following the ideal transformation parameters,
but still have a quasi perfect performance. Meanwhile, the extreme
values exist in the conventional transformation cloak can be
avoided. When applying the genetic algorithms, the thickness of each
layer of the cloak is fixed and the chromosome is a string of 0s and
1s representing a set of constitutive parameters of each layer,
$\{\epsilon_{z1},\mu_{\rho 1},\mu_{\phi 1},...,\epsilon_{z
M},\mu_{\rho M},\mu_{\phi M}\}$. The fitness of an individual is
chosen to be $1/Q_{sca}$, where $Q_{sca}$ is the total scattering of
this individual shown by Eq. (\ref{eq:Qsca}). Making use of the
roulette wheel selection, in which individuals with larger fitness
have larger chance of going forward to the next generation, setting
the single point crossover probability and the mutation probability
to be 0.6 and 0.05, respectively, and ensuring the fittest
individual to propagate to the next generation, evolution is carried
out and optimization is obtained finally. In order to compare the
performance of the cloaks created with the TB parameters and the
optimized parameters, the values of the permittivity and
permeability in the optimization are confined to be positive values
no larger than the maximums of the TB parameters.

\begin{figure}
\includegraphics[width=0.5\columnwidth,draft=false]{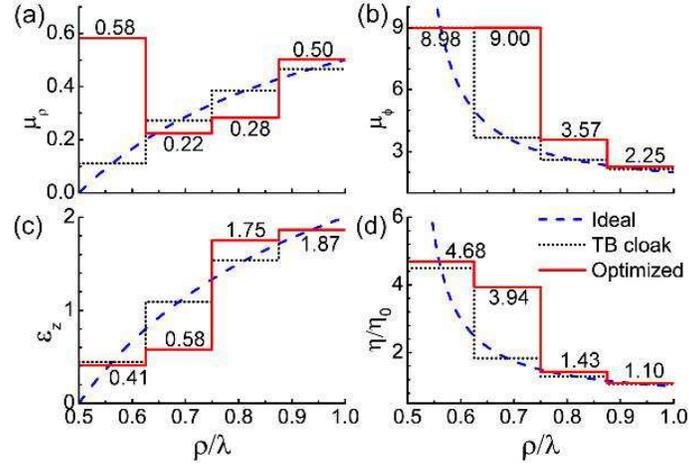}
\caption{\label{fig:parameters} The relative constitutive
parameters and the relative impedance for the 4-layer
quasi-perfect cloak (solid lines), the transformation based cloak
(dotted lines) and the ideal linear cloak (dashed lines).}
\end{figure}

Since in fabrications, it is much easier to create the cloak with
fewer layers of matematerials. We consider the 4-layer cloak as an
example. The dimensions of the cloak are $R_0=\lambda$ and
$R_4=\lambda/2$ with a thickness of $\lambda/8$ for each layer. When
a TE-polarized plane wave normally incident onto a PEC cylinder with
radius $\lambda/2$ the far-field total scattering efficiency
$Q_{sca}$ is found to be 2.31. When the PEC cylinder is coated by
the 4-layer cloak with the TB full parameters, $Q_{sca}$ is reduced
to be 0.18. While using our proposed method, we design a 4-layer
optimized cloak with the parameters shown in
Fig.~\ref{fig:parameters} (a), (b) and (c), the total scattering
efficiency $Q_{sca}$ of this cloak is greatly minimized to be
0.0025.

\begin{figure}
\includegraphics[width=0.5\columnwidth,draft=false]{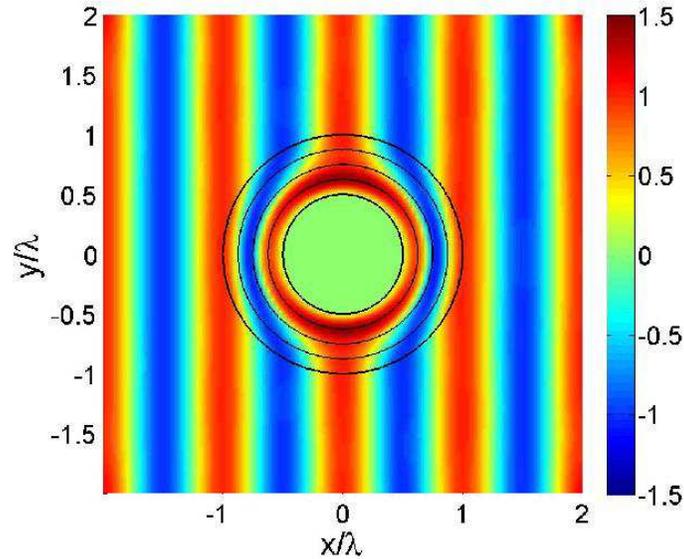}
\caption{\label{fig:optfields} Electric field distributions for a
plane wave incident from left to right onto a quasi-perfect
multilayered cylindrical cloak.}
\end{figure}

\begin{figure}
\includegraphics[width=0.5\columnwidth,draft=false]{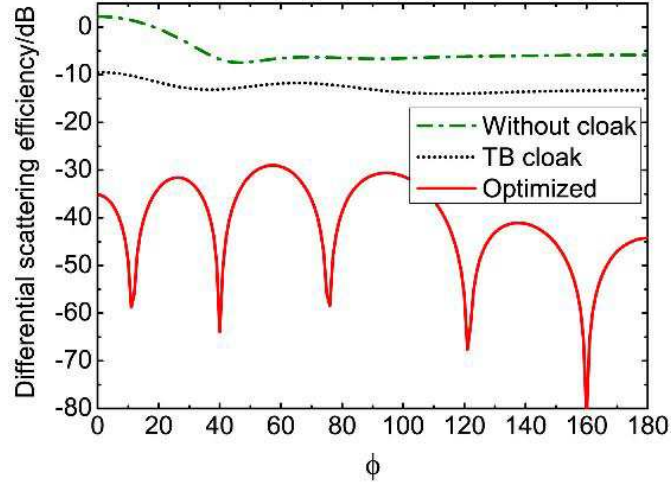}
\caption{\label{fig:optRCS} Comparison of the differential
scattering efficiency of a TE wave incidence for the PEC cylinder
with radius $\lambda/2$ (dotted line), the same PEC cylinder coated
by the TB full parameter multilayered cloak (dashed line) and by the
quasi-perfect cloak (solid line).}
\end{figure}

Fig.~\ref{fig:optfields} shows the calculated electric field
distribution when a TE polarized plane wave normally incident from
left to right onto the optimized cloak. It is shown that in the near
region of the cloak, the electric fields stay almost unperturbed,
thus a so called ``quasi-perfect" cylindrical multilayered cloak is
obtained. In order to see how much improvement has been achieved
through the optimization, the differential scattering efficiency as
a function of the scattering angles is calculated as shown in
Fig.~\ref{fig:optRCS}.
We see the scatterings are minimized by about 10dB when the TB full
parameter 4-layer cloak (dotted line) is used comparing with a bare
PEC core (dash dotted line). When the 4-layer cloak is created with
the optimized parameters (solid line), about 20dB of reduction are
obtained for all the directions comparing with the TB one. Thus a
``quasi-perfect" invisibility cloak is achieved although only 4
layers are used.

The reason why quasi-perfect invisibility can still be achieved in
a cloak with only a few layers can be physically explained as
follows: The scattering of such kind of multilayered cloak is
determined by the recurrence equation Eq. \ref{eq:rto}. When
$m=0$, Eq. \ref{eq:rto} indicates that the $n$th order scattering
is the sum of the direct scattering at $R_0$ denoted by $r_{01n}$
and the wave coming out from $R_0$ caused by the multiple
reflections and transmissions on the inner boundaries, denoted by
$\tilde{t}_{10n}$. In order to minimize the total scattering, the
parameters of the cloak should be chosen so that $r_{01n}$ and
$\tilde{t}_{10n}$ cancel each other. And as denoted by Eq.
\ref{eq:rto}, $\tilde{r}_{01n}$ are actually determined
simultaneously by $\nu_{mn}$, $k_m$ and $\eta_m$ in all the
layers. Thus the match of impedances between the conjoined layers
does not assure a small total scattering, while on the contrary,
the mismatch of the impedances, as shown by
Fig.~\ref{fig:parameters} (d), can be utilized to form multiple
reflections and transmissions among the inner layers, which
eventually produce a transmission to the free space being able to
destructively interfere with the direct reflection occurs at the
outer boundary. In our proposed quasi-perfect cloak, the $0$th
direct reflection coefficient at $R_0$ is
$r_{010}=-0.10741+0.3096i$, while the $0$th transmission
coefficient from $R_0$ is $\tilde{t}_{010}=0.10739-0.3052i$,
therefore, the $0$th scattering coefficient is
$\tilde{r}_{010}=-0.00002+0.0044i$ which is very small, due to the
destructive interference. This is also similar to the
one-dimensional multilayered case, where the impedances of each
layer are not necessary to be matched in order to get zero
reflection.

It's interesting to see from Fig.~\ref{fig:parameters} that the
optimized parameters (solid lines) are quite different from the
parameters obtained by the coordinate transformation (dotted lines).
Taking the $\rho$ component as an example, as shown in
Fig.~\ref{fig:parameters} (a), the value of relative permeability in
the most inner layer is optimized to be 0.58, instead of a
close-to-zero value as suggested by the transformation method. We
also see that the parameters achieved here are relatively small and
within the limit of metamaterials, which shows the possibility of
realizing such a ``quasi-perfect" cloak. This is a very important
contribution to the implementation of the cloak in practice. As we
know, for an ideal cylindrical cloak, the $\phi$ component of the
parameters will go infinity near the inner boundary, as shown by the
dashed line in Fig.~\ref{fig:parameters} (b). Such kind of extreme
value near the inner boundary is very difficult to realize. A
truncation method \cite{ruan2007prl} at the inner boundary can be
used to avoid the extreme value of the inner boundary of the cloak,
however, scattering will be aroused and it has been shown that the
performance of such a cloak is sensitive to the perturbations on the
inner boundary \cite{ruan2007prl}. In order to get a performance as
good as our proposed cloak, here $Q_{sca}$ is 0.0025, only about
$\lambda/10^8$ truncation is allowed on the inner boundary, which
means huge values of permeability about $5\times10^7\mu_0$ is needed
in the $\phi$ direction near the inner boundary, this is a
disadvantage of the transformation based cloak in practical
implementation.

In conclusion, the analytical model for the multilayered cylindrical
cloak has been well established. Based on this model, the effects of
discretization and simplification, as well as losses, of the
transformation based cloak can be quantitatively characterized. By
utilizing the genetic algorithms, we further show that, it is not
the best way to manually assign the transformation parameters for
the multilayered cloak. A 4-layer ``quasi-perfect" cylindrical cloak
is proposed, whose parameters are relatively small, and possible for
realization. The parameters obtained do not follow the method of
coordinate transformation. Our method was shown to be effective in
analyzing a multilayered cloak and provides a robust way of
designing a practical cloak.

This work is supported by the ONR under grant No. N00014-06-01-0001,
the Department of the Air Force under Air Force Contract No.
F19628-00-C-0002, and the NSFC under grant Nos. 60531020 and
60801005.

\bibliographystyle{apsrev}

\end{document}